# RECONSTRUCTION AND ANALYSIS OF CANCER-SPECIFIC GENE REGULATORY NETWORKS FROM GENE EXPRESSION PROFILES

Khalid Raza[1*] and Rajni Jaiswal[2]

[1]Department of Computer Science, Jamia Millia Islamia (Central University),
New Delhi-110025, India.
`kraza@jmi.ac.in`
[2]Department of Computer Science, Jamia Hamdard, New Delhi-110062, India.

## ABSTRACT

*The main goal of Systems Biology research is to reconstruct biological networks for its topological analysis so that reconstructed networks can be used for the identification of various kinds of disease. The availability of high-throughput data generated by microarray experiments fuelled researchers to use whole-genome gene expression profiles to understand cancer and to reconstruct key cancer-specific gene regulatory network. Now, the researchers are taking a keen interest in the development of algorithm for the reconstruction of gene regulatory network from whole genome expression profiles. In this study, a cancer-specific gene regulatory network (prostate cancer) has been constructed using a simple and novel statistics based approach. First, significant genes differentially expressing them self in the disease condition has been identified using a two-stage filtering approach t-test and fold-change measure. Next, regulatory relationships between the identified genes has been computed using Pearson correlation coefficient. The obtained results has been validated with the available databases and literatures. We obtained a cancer-specific regulatory network of 29 genes with a total of 55 regulatory relations in which some of the genes has been identified as hub genes that can act as drug target for the cancer diagnosis.*

## 1. INTRODUCTION

Microarray technology allows researchers to measure the expressions of large numbers (thousands) of genes simultaneously. In human body, all cells contain same genetic material but the same genes may or may not be active. This variation in the activation of genes assists researchers to understand more about the function of the cells. Microarray technology helps researchers to get insight about many different diseases such as various cancer disease, heart disease, mental illness, and infectious disease, etc. [1]. Gene regulation refers to processes in which cells are used to create functional gene products (such as RNA, proteins) from the information stored in genes (DNA). Gene expression data is used widely for the analysis of disease and its diagnosis. Microarray gene expression data is playing a significant role in cancer predication and diagnosis. These data can be characterized by many variables (genes) which are measured on only a few observations (experiments) due to experimental limitations [1]. This provides great opportunities to explore large scale regulatory networks for various purposes such as to identify specific genes causing particular disease so that researchers can target those genes to understand interactions among transcription factors and drug targets, to understand metabolism, and so on [2].

Gene regulatory networks (GRNs) are the systematic biological networks that describe interactions among genes in the form of a graph, where node represents genes and edges their regulatory interactions. Understanding the GRNs helps in understanding interactions among genes, biological and environmental effects and to identify the target genes for drug against the diseases [3]. GRNs have been proved to be a very useful tool used to describe and explain complex dependencies between key developmental transcription factors (TFs), their target genes and regulators [4][5]. Reconstruction of GRN is the development of network model from the available datasets. The GRN reconstruction explicitly represents the developmental or regulatory process, which is of great interest today. Reconstruction has become a challenging computational problem for researchers to understand complex regulatory mechanisms in biological systems. Although, every methods for inferring GRNs from microarray gene expression profiles have both strengths and weaknesses. In this study, we constructed and analysed prostate cancer GRN. Prostate cancer is a slow growing cancer that develops in the prostate and it can spread to other parts of the body such as bones and lymph nodes. It has been reported that prostate cancer is the second leading cause of cancer-related death in United States [6] and sixth in the world [7]. This cancer is most common in developed countries with growing rates in development countries. Monitoring of gene expression from microarray is considered to be one of the most promising techniques for the discovery of GRNs. This technique making GRNs feasible. However, inferring GRNs from time series microarray gene expression involves following challenges i) number of related genes is very large compared to the number of samples or time points, ii) observed data involves a significant amount of noise, and iii) gene interactions displays complex (nonlinear and dynamic) relationships [2,3,8,9].

## 2. RELATED WORKS

The gene regulatory network models can be used to enhance the understanding of gene interactions and explicate the environmental and drug effects. Gene regulatory networks models can mainly be categorized into two types that use discrete and continuous variables [5,10]. The models that use discrete types of variables assume that genes exist in discrete state only. Boolean variables implement these types of approximation in which genes are either in active state (1) or inactive state (0). The Boolean networks are not realistic because some information loss occurs during discretization [11,5]. Bayesian models implements the discretization of variables. These models estimate the probability relationship between genes in the network. The structure of these types of GRNs is modeled by a directed acyclic graph (DAG) in which the expression level, its conditional dependencies on parent and it's probability distribution of particular gene are estimated. These networks are unsuitable for handling time series gene expression or temporal information [5,11,12,13]. The models that uses continuous variables and got most popular are based on ordinary differential equations (ODEs). It models the concentrations of RNAs, proteins, and other molecules with nonnegative real number values of variable. The disadvantage of numerical techniques is the lacking the measurements of the kinetic parameters in the rate equations [5,13]. Most of the earlier work on reconstruction of GRN has been done on smaller organisms having small genome. A few attempt has been made to reconstruction GRN of human related disease. Wang and Gotoh [27] inferred directed cancer-specific GRN using soft computing rules from microarray data. They studied Colon Cancer datasets consisting of 2000 genes and 62 samples and analyzed 18 annotated genes only. Basso et al. [28] report the reconstruction of gene regulatory networks from gene expression profiles of human B cells. The results show a scale-free network, where a few hub genes were identified. Jiang et al [29] identified multiple disease pathways in which genes extracted by supervised learning of the genome-wide transcriptional profiles for patients and normal samples. A pair-wise relevance metric, adjusted frequency value, was applied to describe the degree of genetic relationship between two molecular determinants. The methodology was applied to analyze microarray dataset of colon cancer and results demonstrate that the Colon Cancer-specific gene network captures the most central genetic interactions. The topological analysis of inferred network shows three known hub cancer genes. An extensive review on GRN modelling can be found in [8, 4, 11, 14, 13, 10]. In the present study, we tried to identify hub (highly connected) genes and analyze the topological behaviour of the constructed network in prostate cancer datasets.

## 3. MATERIALS AND METHODS

To find regulatory relationship between gene pairs using gene expression profile, many techniques have been used in the literature. In this work, Pearson's correlation coefficient has been applied. The main steps of the proposed algorithm are outlined as follows.

(1) Preprocessing of the dataset
(2) Identification of most significant genes
(3) Finding regulatory relationship between gene pairs
(4) Elimination of weak correlation
(5) Visualization of the network
(6) Biological validation
(7) Topological analysis

### 2.1. Preprocessing of Datasets

The gene expression data are mostly present in normalized form. The normalized data for each gene are typically known as an 'expression ratio' or as the logarithm of the expression ratio. The expression ratio for a particular gene is basically the normalized value of the expression level which is the ratio of query sample and its normalized value for the control. The datasets under consideration are in normalized form. We did data preprocessing to handle missing values, duplicate and missing gene names, etc in the datasets.

### 2.2. Identification of Most Significant Genes

In this step, those genes are identified that are differentially expressing themselves in diseased condition. A two-stage filtering strategy has been applied in this paper. At the first stage, statistical measure t-test has been applied. The t-test for unpaired data and both for equal and unequal variance can be computed as,

$$t_i = \frac{\bar{y}_i - \bar{x}_i}{\sqrt{\frac{g_i^2}{n_1} + \frac{h_i^2}{n_2}}} \quad (1)$$

where $x_i$ and $y_i$ are the means, $g_i$ and $h_i$ are the variances, and $n_1$ and $n_2$ are the sizes of the two groups of the samples (conditions) tissue and cultured, respectively, of gene expression profile $i$.

At the second stage, a fold-change strategy has been applied. A fold change is a measure that describes how much expression level of a gene changes over two different samples (conditions) or groups. The fold change (FC) for linear data can be calculated as,

$$FC_i = Log_2 \frac{\bar{y}_i}{\bar{x}_i} \text{ or } Log_2 y_i - Log_2 x_i \quad (2)$$

where, $x_i$ and $y_i$ are mean of gene expression profile $i$ in tissue and cultured cases, respectively. In case, gene expression data is already in *Log2* transformed form, fold change can be computed as [15],

27

$$FC_i = \frac{\overline{y_i}}{x_i} \text{ or } y_i - x_i \tag{3}$$

## 2.3. Finding Regulatory Relationship Between Gene Pairs

We applied Pearson correlation coefficient $r_{xy}$ to find out regulatory relationship between gene pairs x and y. The correlation is +1 if there is a perfect positive linear relationship, −1 if there is a perfect negative linear relationship and values between −1 and 1 indicates the degree of linear dependence between the variables. Closer the coefficient to either −1 or +1, stronger the correlation between the variables. If the coefficient is zero, the variables are independent. If we have *n* samples (conditions) of *x* and *y* genes, written as $x_i$ and $y_i$ where *i* = 1, 2, ..., *n*, the correlation coefficient between *x* and *y* ($r_{xy}$) can be estimated as,

$$r_{xy} = \frac{n\sum x_i y_i - \sum x_i \sum y_i}{\sqrt{\left(n\sum x_i^2 - \left(\sum x_i\right)^2\right)\left(n\sum y_i^2 - \left(\sum y_i\right)^2\right)}} \tag{4}$$

Once, the pair-wise correlation coefficient between genes are computed, next we select those coefficient having absolute values above a threshold and eliminated weakly correlated gene pairs. This strategy allows to focus on a few highly connected genes. In this study, we observed that only few genes are strongly correlated, mostly positively and few negatively. The pair-wise correlation among rest of the gene pairs are pretty week.

## 4. RESULTS AND DISCUSSIONS

In this study, microarray data of prostate cancer has been taken for network construction and its topological analysis [30]. The dataset (the full dataset can be downloaded from GEO http://www.ncbi.nlm.nih.gov/geo/query/acc.cgi?acc=GSE26126) consists of 27575 genes, having 181 tissue and 12 cultured samples. To find out the most significant genes, a two-stage filtering has been applied. In the first stage, the statistical test two-tailed t-test is applied to find out the significant genes and considered only those genes having p-value<=0.001 as significant genes and extracted it for the analysis. Out of the 27575 genes from the dataset, 9985 genes have been extracted on the basis of p-value, which is approximately 36% of total number of genes. The formula to calculate t-statistic for unpaired data and both for equal and unequal variance is given in equation (1). In the next stage, we applied a fold-change measure to evaluate the changes in expression level of each gene. The fold change for two kind of data can be calculated using equation (3). At this stage, we considered only those genes showing a minimum of five-fold change in expression-level and finally 101 genes has been selected, which is 0.01 % of 9985 genes (extracted at first stage) and 0.003% of total 27575 genes.

Further, Pearson correlation has been applied and observed the pair-wise correlation among extracted 101 genes. The Pearson correlation can be calculated using equation (4). The week correlation between gene pairs has been dropped. The correlation absolute value which are >=0.85 has been considered as strong correlation and thus, 55 regulatory relationship has been identified which involves 29 genes only. This strategy again reduced the noise level of data up to 0.001% of total of 27575 genes.

Finally, 29 extracted genes has been validated with available biological databases and literatures. Most of the genes among 29 are somehow involved in prostate cancer. The Table 1 shows validation of individual genes and their family from various biological databases and literature. The Table 2 shows the interaction of gene pairs and representing either the relation is activation (+) or repressing (−) .The positive (+) correlation shows activation and negative value (−) shows repressing (inhibiting). Out of 55 extracted regulatory relations, 52 are activators and only 3 are inhibitors.

Table 1. List of genes found to be involved in prostate cancer.

| Genes | Brief description | Reference(s) |
|---|---|---|
| GAS6 family, MYO3A | overexpressed in androgen-independent compared to androgen-dependent prostate cancer cells | A P. Singh, et al. (2008) [16] |
| S100A16 family, PSMD1, RND2, PLAT, PPP3R1family | under-expressed in androgen-independent compared to androgen-dependent prostate cancer cells | A P. Singh, et al. (2008) [16] |
| KCNS2, SLC17A8 family, COL17A1 family | Potential secretory biomarkers for selenium action in prostate cancer | Hongjuan Zhao (2004) [17] |
| PHLDA1 | Changes in Expression in benign prostatic in benign prostatic hyperplasia (BPH) | Hongjuan Zhao, et al (2004) [17] |
| KRT5 | marker of basal cells in prostate glands | Uma R Chandran, et al. (2007) [18] |
| CA9 | gene of C4-2 prostate cancer cell line which is being expressed | Asa J Oudes, et al. (2005) [19] |
| SLAMF9 | SLAMF9 is subfamily of CD2 which in association with EWI subfamily to inversely correlates with metastasis potential of prostate cancer | Xin A. Zhang, et al. (2003) [20] |
| AQP10 | expression and cellular localization of the AQPs were determined in the human prostate cancer | Insang Hwang, et al. (2012) [21] |
| BNIP3 | overexpressed in various tumors, including prostate cancer. | Xueqin Chen, et al. (2010) [22] |
| ZFAND2B | Zinc finger, AN1-type domain 2B expressed in prostate cancer and many more tissues | G2SBC Database [23] |
| SRPX2 | the nucleic acid that encodes a SRPX2 can be act as target for cancer like prostate cancer. | IMHOF, et al. (2007) [24] |
| CSTF1 | 186-gene "invasiveness" gene signature (IGS) including CSTF1 are not only associated with only breast cancer but also in many cancer cells such as prostate cancer, etc. | Rui Liu, (2007) [25] |
| KCNE2 | KCNQ1 form complex with KCNE2 family and regualtes in prostate cancer | NCBI [26] |

At the next step the network has been constructed using Cytoscape software tool which is shown in Fig. 1. From the constructed network in Fig. 1, we can easily identify that genes KRT5, BNIP3, GJB5 and KCNE2 are participating as hub genes with former three having total degree (indegree and outdegree) of eight, and later having total degree of six. GJB5 is activating seven other genes SLAMF9, PAK6, COL17A1, HCAR2, C8A, S100A16 and KRT5, and GJB5 is activated by other hub gene KCNE2. Similarly, gene KCNE2 activates six other genes GJB5, SLAMF9, COL17A1, HCAR2, PAK6 and KRT5. The gene CSTF1 does not activate any other gene rather it is activated by EMP1 and inhibited by two other genes SRPX2 and hypothetical LOC401459. The hypothetical LOC401459 inhibits only CSTF1, activated by SRPX2 and inhibited by EMP1. There are many genes that do not regulate (either activate or inhibit) any other genes in the network such as PSMD1, CSTF1, ZFAND2B, BNIP3, PLAT, C8A and HCAR2. From the Fig. 1 it is clear that gene BNIP3 is activated by large number of genes and hence it will be overexpressed. From the literature [22] it has been proved that gene BNIP3 is overexpressed in various tumors including prostate. The other identified hub gene KRT5 is a marker of basal cells in prostate glands and shows uniform downregulation in all metastatic tumors [18]. From the literature [26], it has been validated that KCNQ1 form complex with KCNE2 (also one of the identified hub gene) family and regulates in prostate cancer.

Table 2. List of genes found to be involved in prostate cancer.

| Source | Target | Activate (+) Repress (−) |
|---|---|---|
| GJB5 | SLAMF9 | + |
| | PAK6 | + |
| | COL17A1 | + |
| | HCAR2 | + |
| | C8A | + |
| | S100A16 | + |
| | KRT5 | + |
| KCNE2 | GJB5 | + |
| | SLAMF9 | + |
| | COL17A1 | + |
| | HCAR2 | + |
| | PAK6 | + |
| | KRT5 | + |
| S100A16 | PAK6 | + |
| | PHLDA1 | + |
| | C8A | + |
| | PLAT | + |
| ZNF577 | RND2 | + |
| | SLC17A8 | + |
| | GAS6 | + |
| | BNIP3 | + |
| GAS6 | RND2 | + |
| | BNIP3 | + |
| | SLC17A8 | + |
| | PAK6 | + |
| KRT5 | PAK6 | + |
| | BNIP3 | + |
| | ST18 | + |
| COL17A1 | BNIP3 | + |
| | C8A | + |
| | HCAR2 | + |
| AQP10 | SLC17A8 | + |
| | ST18 | + |
| | RND2 | + |
| EMP1 | SRPX2 | + |
| | CSTF1 | + |
| | hypothetical LOC401459 | − |
| PPP3R1 | ZFAND2B | + |
| | YWHAH | + |
| CA9 | S100A16 | + |
| | AQP10 | + |
| SRPX2 | hypothetical LOC401459 | + |
| | CSTF1 | − |
| SLC17A8 | RND2 | + |
| | BNIP3 | + |

30

| PAK6 | BNIP3 | + |
| --- | --- | --- |
|  | RND2 | + |
| RND2 | BNIP3 | + |
| PHLDA1 | PLAT | + |
| YWHAH | PSMD1 | + |
| KCNS2 | SLC17A8 | + |
| SLAMF9 | HCAR2 | + |
| MYO3A | RND2 | + |
| C8A | PAK6 | + |
| hypothetical LOC401459 | CSTF1 | − |

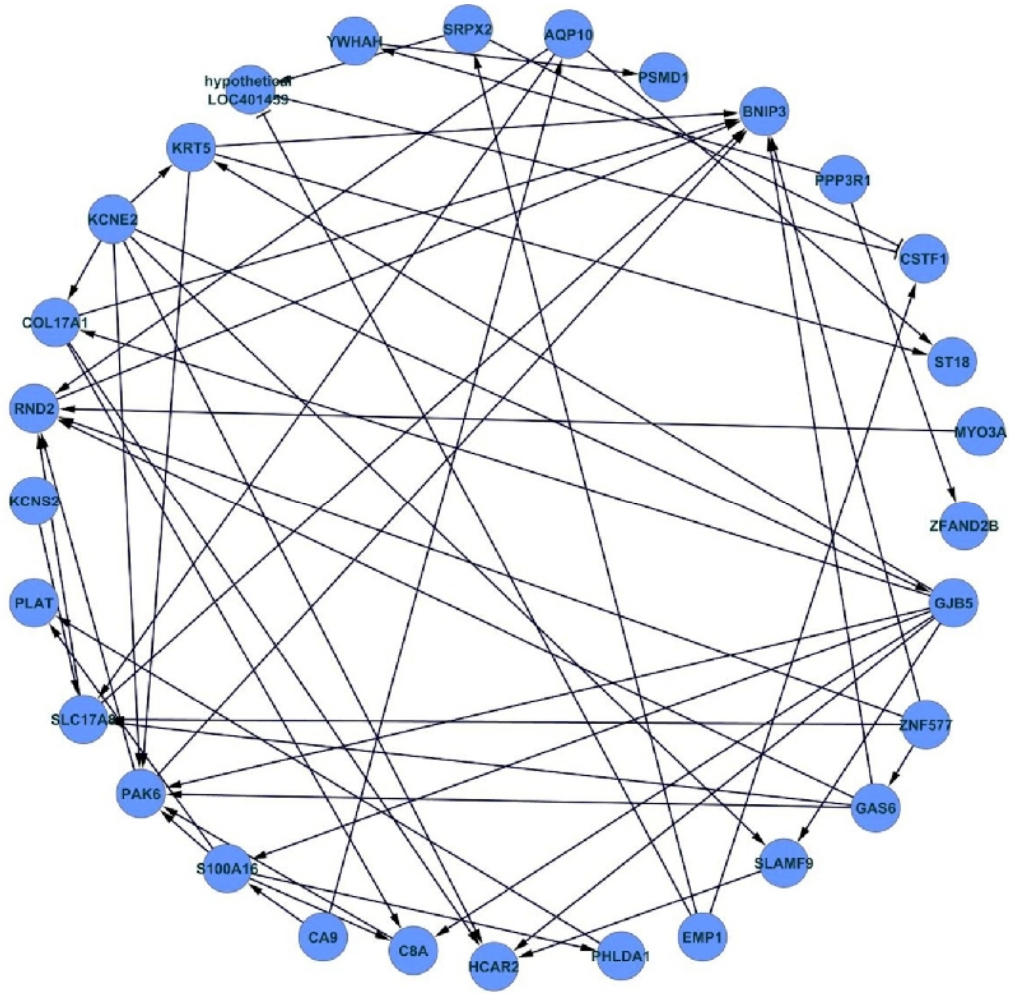

Figure 1. Inferred gene regulatory network of 29 genes and 55 regulatory relations using proposed methodology. The finding shows that genes KRT5, BNIP3, GJB5 and KCNE2 are participating as hub genes.

International Journal on Bioinformatics & Biosciences (IJBB) Vol.3, No.2, June 2013## 5. CONCLUSIONS

The complex molecular interactions underlying cancer is due to the perturbations in the gene regulatory networks. Therefore, identification of cancerous genes, pathways control by them through gene regulatory networks is a key step towards cancer diagnosis. A directed regulatory network is proficient to reveal interactions among genes more legitimately and also capable to capture cause-effect relations between genes-pairs. This paper reports a simple statistical approach to extract differentially expressed genes, finding correlations between gene-pairs for the reconstruction of gene regulatory networks under specific disease conditions that assist the interpretability of the network. First, genes relevant to a specific cancer using a t-test and fold-change method has been identified. The pair-wise correlation coefficients among gene pairs were calculated and a threshold value has been imposed to eliminate weakly correlated gene pairs and found 55 significantly correlated gene pairs that involves 29 genes. A regulatory network has been constructed using Cytoscape software tool. During the analysis of the constructed network we observed that some genes are working as hub genes including KRT5, BNIP3, GJB5 and KCNE2. Among them, BNIP3 is highly activated (overexpressed) gene which has been proved to be overexpressed in prostate cancer [22]. The other hub gene KRT5 is a marker of basal cells in prostate glands and shows uniform downregulation in all metastatic tumors [18]. The result shows that gene KCNE2 regulate large number of genes which can be validated with [26] that it regulates in prostate cancer.

The regulatory relationships among genes in cancer are not freely accessible from database and available in literature. Due to this problem, the construction of gene regulatory networks and their validation in a realistic manner is really a difficult task. The utility and reliability of our study needs further experimental validation. Our finding can help to reveal common molecular interactions in the cancer under study and provide new insights in cancer diagnostics, prognostics and therapy. Our proposed approach can also be used to investigate other disease specific gene regulatory network like colon cancer, lung cancer, breast cancer and so on. In future study, we will try to construct regulatory networks for other types of cancer from microarray data. Microarray data are inherently noisy due to experimental limitations. Noises in the dataset directly reflects the statistical techniques. Today, artificial intelligence based approach such as fuzzy logic, neural networks, evolutionary computation are being used in many bioinformatics research problems. The promises of fuzzy logic to tolerate noise and deal with impression, neural network to learn from data rich environment and evolutionary computation for the optimization can be good candidate to infer gene regulatory network from microarray data. In the future, we can apply these artificial intelligence based sophisticated techniques to better construct cancer-specific regulatory networks.

## ACKNOWLEDGEMENTS

The authors would like to thank all scientists behind the publicly available data sets. The author K. Raza acknowledges the funding from University Grants Commission, Govt. of India through research grant 42-1019/2013(SR). The co-author R. Jaiswal acknowledges the Department of Computer Science, Jamia Millia Islamia, New Delhi, India for providing necessary facilities to carry out this research.## REFERENCES

[1]  K.Vaishali & A.Vinayababu, (2011). "Application of microarray technology and softcomputing in cancer biology : a review", *International Journal of Biometrics and Bioinformatics (IJBB)*, vol. 5, no. 4. pp. 225-233.

[2]  Jeffrey D. Allen, et al., (2012). "Comparing statistical methods for constructing large scale gene networks", *PLoS ONE*, vol. 7, no. 1, pp. e29348.32